# Emergence of Layer Stacking Disorder in *c*-axis Confined MoTe$_2$

James L Hart[1], Lopa Bhatt[2], Yanbing Zhu[3], Myung-Geun Han[4], Elisabeth Bianco[2], Shunran Li[5], David J Hynek[6,7], John A Schneeloch[8], Yu Tao[8], Despina Louca[8], Peijun Guo[5,6], Yimei Zhu[4], Felipe Jornada[3], Evan J Reed[3], Lena F Kourkoutis[2,9], Judy J Cha[1,6,7]

1. Department of Materials Science and Engineering, Cornell University, United States
2. Applied and Engineering Physics, Cornell University, United States
3. Department of Materials Science and Engineering, Stanford University, United States
4. Condensed Matter Physics and Materials Science Department, Brookhaven National Laboratory, United States
5. Department of Chemical and Environmental Engineering, Yale University, United States
6. Energy Sciences Institute, Yale University, United States
7. Department of Mechanical Engineering and Materials Science, Yale University, United States
8. Department of Physics, University of Virginia, United States
9. Kavli Institute at Cornell for Nanoscale Science, Cornell University, United States

The layer stacking order in 2D materials strongly affects functional properties and holds promise for next generation electronic devices. In bulk, octahedral MoTe$_2$ possesses two stacking arrangements, the Weyl semimetal T$_d$ phase, and the higher-order topological insulator 1T´ phase; however, it remains unclear if thin exfoliated flakes of MoTe$_2$ follow the T$_d$, 1T´, or an alternative stacking sequence. Here, we resolve this debate using atomic-resolution imaging within the transmission electron microscope. We find that the layer stacking in thin flakes of MoTe$_2$ is highly disordered and pseudo-random, which we attribute to intrinsic confinement effects. Conversely, WTe$_2$, which is isostructural and isoelectronic to MoTe$_2$, displays ordered stacking even for thin exfoliated flakes. Our results are important for understanding the quantum properties of MoTe$_2$ devices, and suggest that thickness may be used to alter the layer stacking in other 2D materials.

In layered van der Waals (vdW) solids, exotic quantum phenomena can be engineered *via* the layer stacking. For instance, the twist angle in bilayer graphene influences the low-energy electronic band structure, allowing for the control over magnetic[1], superconducting[2], and topological phases[3]. When the twist angle of a homo-structure is zero, the in-plane displacement between layers, *i.e.* the layer stacking order, offers an additional control parameter. Examples include emergent ferroelectricity in hexagonal boron nitride[4,5], magnetic order in CrI$_3$[6], and quantum transport in trilayer graphene[7]. In certain cases, the layer stacking order can be dynamically controlled through external stimuli[8–10], which is attractive for device applications. However, our basic understanding of layer stacking energetics, as well as layer sliding transitions, remains limited.

Octahedrally coordinated MoTe$_2$ and WTe$_2$ are prime candidates for stacking order-dependent devices. In bulk, two stable stacking arrangements exist[11,12]: the low temperature T$_d$ phase, a ferroelectric Weyl semimetal[13], and the high temperature 1T´ phase, a higher order topological insulator[14] (Fig. 1a) with the transition temperature ($T_c$) of ~250 K for MoTe$_2$[15] and ~565 K for WTe$_2$[16]. In thin mechanically-exfoliated flakes of MoTe$_2$, the temperature-dependent layer stacking transition is suppressed; however, the preferred layer stacking in such flakes is debated. Raman spectroscopy studies have reached conflicting conclusions, finding either that thin flakes prefer 1T´ stacking, or T$_d$ stacking, or alternative stacking sequences distinct from the known bulk phases[17–22]. In this thickness range, MoTe$_2$ flakes show myriad intriguing phenomena, *e.g.* enhanced superconductivity[18], superconducting edge currents[23], giant out-of-plane Hall effect[24], and in-plane, third-order nonlinear Hall effect[25]. To fully understand and exploit these behaviors, the stacking order – which dictates the symmetry and topology – must be determined. Moreover, determining the stacking in thin MoTe$_2$ may serve as a general platform for understanding dimensional effects in other 2D materials. In contrast to MoTe$_2$, the thickness-dependence of layer stacking in WTe$_2$ has not been studied.

Here, we determine the structure of MoTe$_2$ and WTe$_2$ flakes by atomic-resolution scanning transmission electron microscopy (STEM) imaging. We find that the layer stacking in thin exfoliated flakes of MoTe$_2$ does not follow the ordered 1T´ or T$_d$ phases, rather, the stacking is highly disordered. In contrast, the stacking in WTe$_2$ is well ordered T$_d$, even for thin flakes. To explain the disordered stacking in MoTe$_2$, we exclude extrinsic mechanisms such as sample oxidation and interface effects, and discuss intrinsic coupling between flake thickness, stacking arrangement, and free energy. These results are crucial for our interpretation of the various quantum properties exhibited by MoTe$_2$ flakes, and for the future design of MoTe$_2$ based devices.

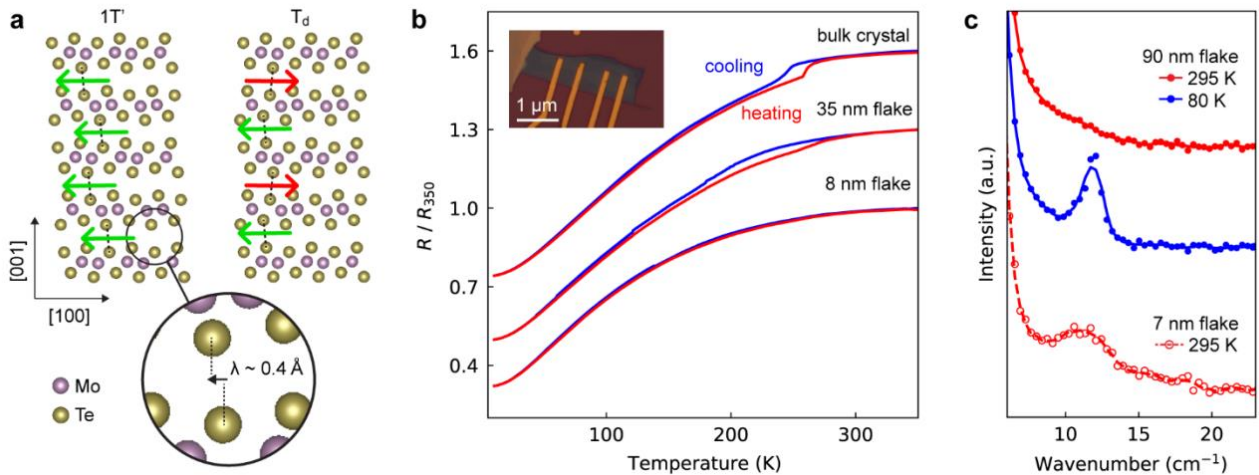

**Figure 1. a.** Schematics of the 1T´ and T$_d$ phases of MoTe$_2$ viewed down the [010] zone axis. The overlaid arrows represent the in-plane component of the Te-Te pairs that bridge the vdW gap, here expressed by the vector λ. The color and length of the arrows denote the displacement direction and amplitude of λ. **b.** Electrical resistance of MoTe$_2$ of varying thicknesses as a function of temperature, normalized to the resistance at 350 K. Data are offset vertically for clarity. Inset shows an optical image of the 8 nm flake. **c.** Raman spectroscopy of MoTe$_2$ flakes as a function of temperature and flake thickness. The spectra show the inter-layer shear mode, which is sensitive to the layer stacking order[19].

**Results**

*Electrical transport and Raman spectroscopy:*

We initially studied the layer stacking phase transition of MoTe$_2$ through electrical transport measurements and Raman spectroscopy. Transport measurements of a bulk crystal show a clear thermal hysteresis loop centered at ~250 K, indicative of the first-order stacking transition (Fig. 1b)[15]. The thermal hysteresis loop is broadened and partially suppressed for flakes 10s of nm thick, and then fully suppressed for flakes < 10 nm. This trend suggests that the stacking transition is mostly quenched in thin exfoliated flakes, consistent with prior reports[17,18,22].

With Raman spectroscopy, the most direct signature of the stacking transition in bulk MoTe$_2$ is the activation of an inter-layer shear mode at 12 cm$^{-1}$ (1.5 meV)[19]. This mode is Raman silent in the centrosymmetric 1T´ phase, but emerges in the T$_d$ phase owing to inversion symmetry breaking[26,27]. For our measurements of a 90 nm thick flake, the inter-layer shear mode is absent at room temperature as expected, and then activated at 80 K, consistent with the bulk 1T´ to T$_d$ transition (Fig. 1c). Conversely, for a 7 nm thick flake, we unexpectedly observe the shear mode at room temperature (50 K above the bulk $T_c$), though the peak is broadened and softened. This finding is similar to that of ref. [17].

Taken together, our transport and Raman data suggest that for thin flakes, the stacking transition is quenched, and the T$_d$ phase is stabilized up to (at least) room temperature. This interpretation has been advocated in prior reports[17,18]. However, the precise relation between the layer stacking and the electrical resistance is unclear[28,29]. Moreover, the emergence of the inter-layer Raman mode does not guarantee the T$_d$ phase, rather, this mode simply indicates inversion symmetry breaking[26,27]. Alternative stacking sequences could also break inversion symmetry, and for thin flakes, inversion symmetry is necessarily broken at interfaces even for centrosymmetric crystals. Hence, symmetry-based Raman analysis cannot unequivocally identify the layer stacking, and direct atomic-scale visualization is needed.

*Room temperature (S)TEM:*

To directly determine the structure of thin exfoliated MoTe$_2$ and WTe$_2$ flakes, we performed STEM high angle annular dark field (HAADF) imaging. First, we observe flakes in plan-view (the *ab*-plane) by transferring exfoliated flakes to a STEM grid *via* a PDMS stamp. In this geometry, the 1T´ and T$_d$ stacking sequences are easily differentiated, as shown with the atomic schematics and STEM simulations in Fig. 2a. Our experimental imaging of exfoliated MoTe$_2$ flakes reveal several distinct structures, none of which perfectly match the 1T´ or T$_d$ phases (Fig. 2b). These results indicate that the layer stacking in thin MoTe$_2$ flakes is not spatially uniform, and that the stacking does not follow either of the bulk phases. In contrast, for exfoliated WTe$_2$, the observed crystal structure is in good agreement with the simulated T$_d$ structure, demonstrating that thin WTe$_2$ flakes possess ordered T$_d$ stacking at room temperature, the same as bulk WTe$_2$ (Fig. 2c).

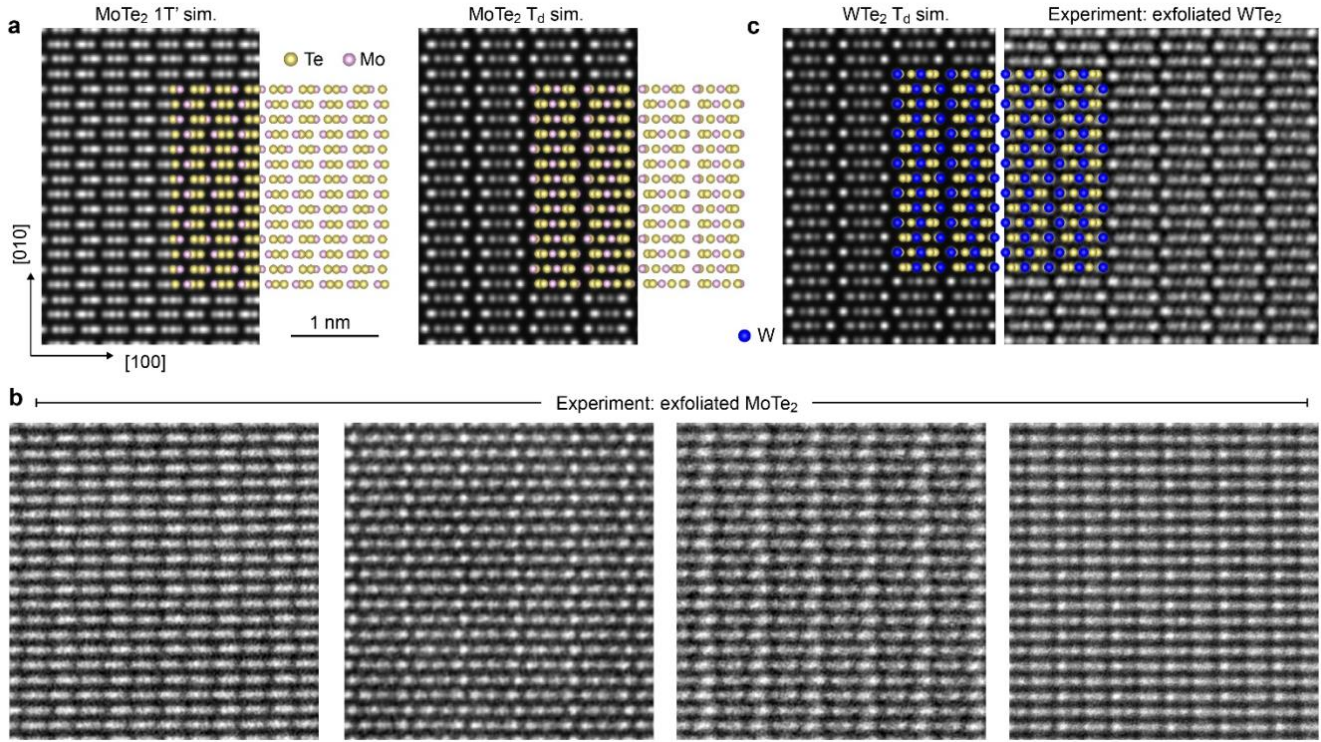

**Figure 2. a.** Simulated STEM-HAADF images of 1T´ and $T_d$ MoTe$_2$, with overlaid atomic schematics. **b.** Several experimental STEM-HAADF images of exfoliated MoTe$_2$ in plan-view (*ab*-plane), none of which match the simulated $T_d$ or 1T´ images. **c.** Simulated and experimental STEM data for exfoliated WTe$_2$. The experimental data matches the $T_d$ simulation. The scale bar in **a** applies to **b** and **c** as well.

To better understand the irregular stacking in MoTe$_2$, we next studied flakes in cross-section (the *ac*-plane), which allows direct determination of the layer stacking order. As schematically shown in Fig. 1a, the 1T´ and $T_d$ phases can be differentiated based on the Te-Te pairs bridging the vdW gap. For 1T´ stacking, the in-plane component of this pair (the inter-layer shift) is always in the same direction (↓↓↓↓ or ↑↑↑↑), while for the $T_d$ phase, the shift direction alternates (↑↓↑↓). Figure 3a shows a STEM image of a bulk MoTe$_2$ crystal, prepared in the cross-sectional geometry using a focused ion beam (FIB). The inter-layer shift (referred to as λ in Fig 1a) is clearly visible. To quantify the inter-layer shift, we fit all Te columns with a 2D Gaussian and directly calculate λ for each Te-Te pair (Supplementary Note 1, Supplementary Figure 1). We then average λ laterally across the image width for each layer. Figures 3b and 3c demonstrate this method on bulk crystals of MoTe$_2$ and WTe$_2$, respectively. The bulk MoTe$_2$ structure follows the expected room temperature 1T´ stacking sequence (↓↓↓↓), and bulk WTe$_2$ follows the expected $T_d$ stacking sequence (↑↓↑↓). We note that both bulk crystals show some degree of stacking disorder, *e.g.* twin boundaries (Supplementary Figure 2).

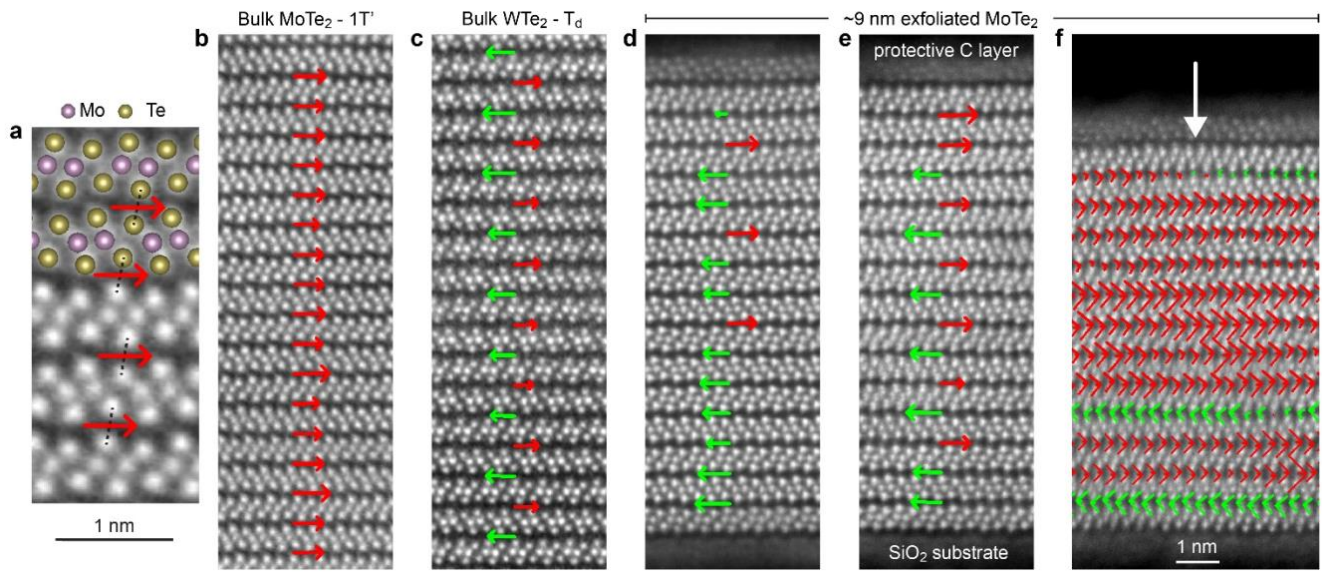

**Figure 3. a.** Magnified STEM-HAADF image of bulk MoTe$_2$ with an overlaid atomic structure schematic. The dotted black lines show the bridging Te-Te pairs, and the red arrows represent the inter-layer shift. **b, c.** STEM-HAADF images of bulk MoTe$_2$ and WTe$_2$, respectively, with overlaid arrows representing the calculated value of λ. **d, e.** STEM-HAADF images of a MoTe$_2$ flake exfoliated onto amorphous SiO$_2$, with a protective carbon overlayer. The arrow magnitudes in **b** - **e** are 15 times the calculated shift. **f.** STEM-HAADF image of the same MoTe$_2$ flake, highlighting a stacking soliton (marked with the white arrow). λ is shown for each individual Te-Te pair. The scale bar in **f** applies to **b** – **e** as well. For **a-f**, samples were viewed down the [010] zone axis. Analysis down the [100] zone axis is shown in Supplementary Figure 1.

We next examine STEM data from a ~9 nm thick MoTe$_2$ flake exfoliated onto a SiO$_2$ / Si substrate. Figure 3d-f shows different regions from the same flake, with the regions separated laterally by 100s of nm. In stark contrast to bulk MoTe$_2$ which shows ordered 1T´ stacking, the thin flake of MoTe$_2$ displays an array of alternative stacking arrangements. There is no strong preference for either 1T´ or T$_d$ stacking, and instead, the stacking is best described as disordered. Changes in layer stacking are accommodated at stacking solitons[30], where the inter-layer shift changes direction. An example soliton is highlighted in Fig. 3f, where arrows representing λ for each individual Te-Te pair are shown. The disordered layer stacking explains the inter-layer Raman mode observed in thin MoTe$_2$ flakes at room temperature (see our data in Fig. 1c and refs. [17] and [19]), since disordered stacking can locally break inversion symmetry. Disordered stacking also explains the multitude of crystal structures observed with plan-view imaging (Fig. 2b).

To characterize the layer stacking on a more global scale, rather than small regions examined by STEM, we performed electron diffraction measurements using a ~3 micron selected-area aperture. We focus on scattering along (2, 0, $L$), which allows easy differentiation between the various stacking geometries. Specifically, simulations show that T$_d$ stacking yields diffraction spots at $L = N$ (N is any integer), 1T´ gives spots at $L = N \pm \delta$ (δ is related to the β angle of 1T´, and the spot doubling is due to twinning of ↓↓↓↓ and ↑↑↑↑ domains), and disordered stacking results in diffuse scattering along $L$[31] (Fig. 4a). For the disordered simulation, we constructed a 'random stacking model' by fixing the magnitude of λ at 45 pm, and randomly selecting the shift direction for each new layer (Supplementary Note 2).

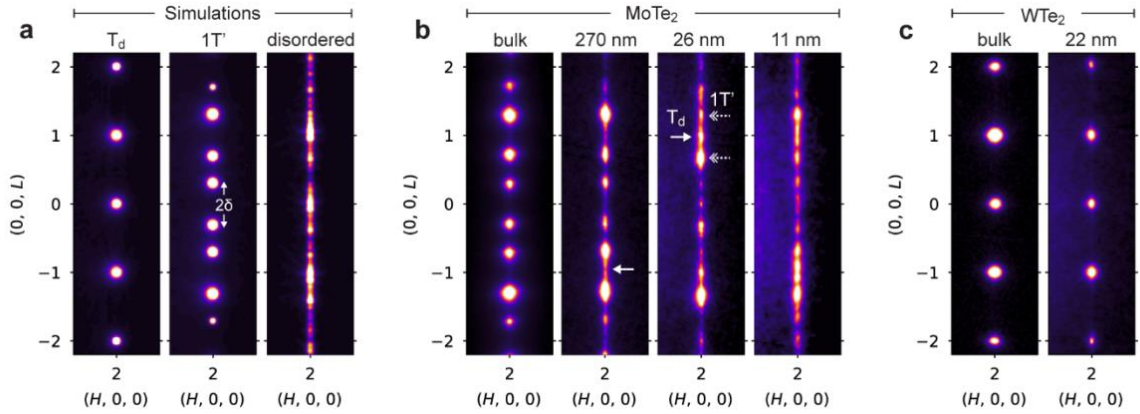

**Figure 4. a.** Electron diffraction simulations of MoTe$_2$ with 1T´, T$_d$, and random stacking. **b.** Experimental TEM selected-area diffraction of MoTe$_2$ as a function of flake thickness. For the 270 nm flake, the arrow indicates diffuse scattering along $L$. For the 26 nm flake, the arrows highlight both 1T´ and T$_d$ diffraction spots. **c.** Experimental data for bulk and thin-flake WTe$_2$. For **a-c**, the diffraction data is indexed using the orthorhombic T$_d$ unit cell.

Experimentally, diffraction from bulk MoTe$_2$ matches the 1T´ simulation, as expected (Fig. 4b). For a 270 nm thick MoTe$_2$ flake, we observed 1T´ diffraction spots; however, diffuse scattering along $L$ is also present (see arrow), indicating a measurable degree of stacking disorder. For flakes 26 and 11 nm thick, spots corresponding to both 1T´ and T$_d$ are present (see arrows), and the diffuse scattering along $L$ is further enhanced. These diffraction measurements show that stacking disorder in MoTe$_2$ flakes is a global effect. In contrast to MoTe$_2$, both bulk and exfoliated WTe$_2$ show ordered T$_d$ stacking, with no apparent thickness effect (Fig 4c).

*Cryogenic electron diffraction:*

Having established disordered stacking in thin MoTe$_2$ flakes at room temperature, but not in thin WTe$_2$ flakes, we next consider whether well-ordered T$_d$ stacking can be stabilized in MoTe$_2$ at sufficiently low temperature. We find that FIB sample preparation restricts layer sliding in bulk MoTe$_2$ (Supplementary Figure 3), thus our approach of cross-sectional (S)TEM analysis cannot be used to study temperature effects. Instead, flakes must be studied in plan-view[32]. Unfortunately, there is no direct method to determine the layer stacking in this geometry: analysis of plan-view STEM-HAADF imaging is complicated owing to the stacking disorder, and the $(H, K, 0)$ diffraction pattern symmetry is completely insensitive to the layer stacking. However, the layer stacking influences the $(H, K, 0)$ diffraction spot *intensities*, which we use to infer the layer stacking.

Figure 5a shows *ab*-plane diffraction data for WTe$_2$ at room temperature, as well as MoTe$_2$ at both room temperature and ~17 K. From electron energy loss spectroscopy analysis, the WTe$_2$ flake is ~25 nm thick, and the MoTe$_2$ flake is ~35 nm thick. We first analyze the data qualitatively. Starting with WTe$_2$, we observe a clear first order Laue zone, which we highlight in Fig. 5b by plotting the intensity of each diffraction spot as a function of momentum transfer, $Q$. The Laue zone is indicative of out-of-plane order with a real-space periodicity of 14 Å, in good agreement with the WTe$_2$ T$_d$ $c$-lattice parameter. In contrast, for the MoTe$_2$ flake measured at room temperature, there is no Laue zone. This indicates a lack of out-of-plane order, consistent with pseudo-random layer stacking. This finding is in agreement with our cross-section (S)TEM analysis shown in Figs. 3 and 4. Upon cooling the MoTe$_2$ flake using liquid He TEM, a weak Laue zone emerges, suggesting the presence of partially ordered layer stacking (Fig. 5b).

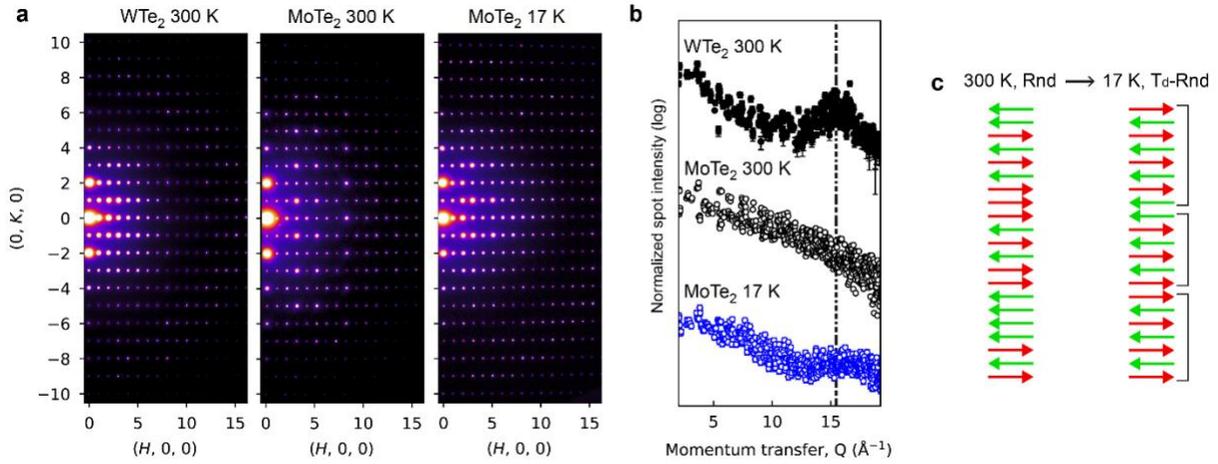

**Figure 5. a.** Experimental TEM diffraction data for WTe$_2$ and MoTe$_2$ flakes taken at room temperature and at ~17 K. **b.** Spot intensity as a function of momentum transfer, $Q$, for WTe$_2$ and MoTe$_2$. The vertical dashed line marks the expected 1$^{st}$ order Laue Zone for both 1T´ and T$_d$ phases. **c.** Schematic of a random stacking sequence (Rnd) versus our T$_d$-Rnd model, which lacks long-range order, but possess short-range T$_d$ order. The brackets highlight local T$_d$ domains.

We next analyze the diffraction data quantitatively. To do so, we extract all the experimental ($H$, $K$, 0) diffraction spot intensities with $H \leq 5$ and $K \leq 3$. We then model the spot intensities with multi-slice electron scattering calculations, using the flake thickness, orientation, and bending as fitting parameters. Fits are performed assuming T$_d$, 1T´, and disordered stacking, and the resulting $\chi^2$ values are compared (Supplementary Note 2). The results are outlined in Table 1. For WTe$_2$ measured at room temperature, our quantitative fitting method strongly favors T$_d$ stacking. This finding is consistent with our atomically-resolved STEM imaging (Fig. 2c), and demonstrates the validity of our quantitative diffraction approach. For the MoTe$_2$ flake measured at room temperature, the random stacking model provides an excellent fit, consistent with the cross-sectional (S)TEM data (Figs. 3 and 4). Upon cooling to ~17 K, the data is best fit with a model that lacks long-range stacking order, but favors local T$_d$ stacking, with an average T$_d$ domain thickness of ~6 layers (4 nm). This model, labeled T$_d$-Rnd, is schematically illustrated in Fig. 5c. We conclude that there is an increase in the relative fraction of T$_d$ stacking after cooling to liquid-He temperatures, but there is no transition to a fully ordered T$_d$ state for MoTe$_2$. This conclusion is supported by both our qualitative Laue zone analysis, and the quantitative multi-slice fitting method.

**Table 1.** $\chi^2$ fitting results from our quantitative diffraction approach. For each row, the bold value represents the best fit. The listed temperatures are approximate. Further details given in Supplementary Note 2 and Supplementary Figure 4

|  | Temp. (K) | T$_d$ | 1T´ | Random | T$_d$-Rnd |
|---|---|---|---|---|---|
| WTe$_2$ | 300 | **3.8** | 40 | 50 | - |
| MoTe$_2$ | 300 | 18 | 15 | **4.3** | - |
| MoTe$_2$ | 17 | 9.2 | 14 | 4.8 | **2.4** |

We also performed *in situ* annealing of MoTe$_2$ with plan-view electron diffraction. With heating up to 675 K, there was no emergence of a Laue zone, and changes in the diffraction spot intensities were within the measurement error (Supplementary Figure 5). Thus, even with high temperature heating, there is no transition to ordered 1T´, and no measured changes in the layer stacking.

*Temperature-thickness phase diagram:*

Compiling all our experimental data, we construct an approximate temperature-thickness phase diagram for MoTe$_2$ (Fig. 6 and Supplementary Note 3). For flakes ≤ 10 nm in thickness, the stacking is highly disordered with minimal thermal dependence. For flakes 10s of nm in thickness, the room temperature stacking is also mixed and disordered. Upon cooling flakes in this thickness range, there is an increase in local T$_d$ order, but the transition is only partial, and disorder persists down to the lowest measured temperatures. For thicknesses at and above ~100 nm, there is a well-defined thermally-driven 1T´ to T$_d$ transition, albeit with increased disorder relative to the bulk. We emphasize that with reduced flake thickness, the effect is not the stabilization of either the 1T´ or T$_d$ structures, rather, with reduced thickness we observe a transition from ordered to disordered layer stacking. Lastly, we note that in the monolayer and bilayer limits, there is no distinction between 1T´ and T$_d$ stacking.

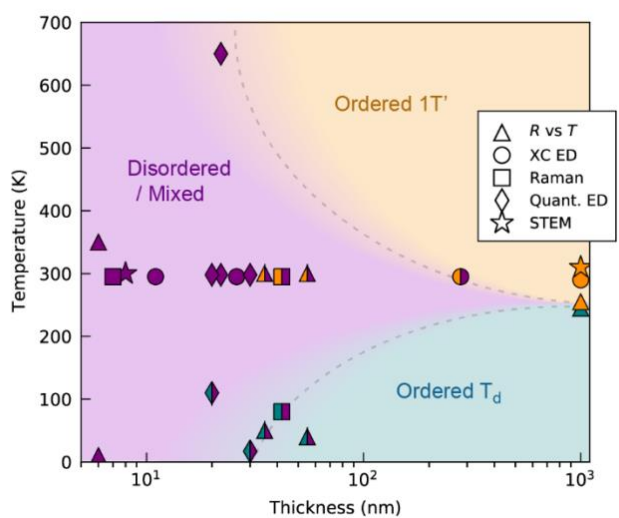

**Figure 6.** Qualitative phase diagram for octahedral MoTe$_2$ as a function of temperature and flake thickness. For each experimental datapoint, the marker shape represents the experiment type, and the color represents the phase. *R* vs *T* indicates electrical resistance measurements (Fig. 1b), XC ED stands for cross-sectional electron diffraction (Fig. 4), and Quant. ED stands for quantitative electron diffraction (Fig. 5). 1T´ stacking is shown in orange, T$_d$ in teal, and disordered / mixed stacking in purple. Note that the emergence of disordered stacking is gradual, and the boundaries separating the 1T´, T$_d$, and disordered phases are purposefully blurred. Further details are given in Supplementary Note 3.

*Consideration of extrinsic effects for stacking disorder:*

We now consider possible origins of the stacking disorder, beginning with sample oxidation. The flakes discussed in Figs. 2 – 5 were exposed to ambient atmosphere prior to (S)TEM analysis, thus oxidation effects may be present. However, we also examined a flake capped with graphene in an Ar glovebox, and observed similar layer stacking disorder (Supplementary Figure 6). Additionally, we imaged the top surface of a bulk crystal exposed to atmosphere (Supplementary Figure 7), and we observed well-ordered 1T´ stacking up to the topmost layers. Hence, the observed stacking disorder is not due to oxidation. Additionally, the observation of well-ordered 1T´ stacking at the surface of a bulk crystal (Supplementary Figure 7) eliminates the possibility of inherent surface effects, *e.g.* the breaking of [001] translational symmetry. Interfacial effects from the $SiO_2$ substrate can also be ruled out, since free-standing $MoTe_2$ flakes show stacking disorder as well (Figs. 2 and 5).

The starting crystal quality, such as the concentration of vacancies and interstitials, may influence the layer stacking. However, our bulk $MoTe_2$ crystal displayed well-ordered 1T´ stacking with large (> 40 layer thick) twin domains (Supplementary Figure 2), as well as a sharp first order phase transition in electrical resistance (Fig. 1b). Moreover, three separate sources of bulk $MoTe_2$ were tested, and all displayed disordered stacking for thin flakes (Materials and Methods). Hence, the parent crystal quality cannot explain the disordered stacking. Electron irradiation effects must also be considered[32]. In our STEM cross-section experiments, we observed the gradual amorphization of $MoTe_2$ and $WTe_2$ in a layer-by-layer fashion; however, we did not observe any layer sliding under the electron beam. Thus, STEM-induced effects cannot account for the observed disorder.

Scotch tape exfoliation (and the associated mechanical strain) offers a possible explanation for our findings; however, there are several reasons to doubt this hypothesis. First, with our experimental method, the normal and shear strains applied to a flake during exfoliation are not a function of the flake thickness (Supplementary Note 4). Thus, this hypothesis suggests that all exfoliated flakes, regardless of thickness, should show the same level of disorder. Conversely, we find that the disorder is greatly enhanced for thin flakes (Fig. 4). Secondly, the inter-layer force constant – which would resist any mechanically-induced layer sliding – is comparable for $WTe_2$ and $MoTe_2$[19,33]. If mechanical exfoliation were responsible for the disordered stacking in $MoTe_2$, then exfoliated $WTe_2$ should show similar levels of disorder, which is not the case (Fig. 2 and 4). Mechanical strain might also cause disordered shifts also along the [010] axis in $MoTe_2$, but this is not observed experimentally (Supplementary Figure 1). Finally, if the observed disorder were simply due to mechanical strain, then one might expect a high temperature anneal to restore the equilibrium, *i.e.* ordered, stacking arrangement. Instead, we find minimal changes in layer stacking after annealing flakes up to 675 K (Supplementary Figure 5).

To summarize, the observed layer stacking disorder in thin $MoTe_2$ flakes cannot be attributed to oxidation, interfacial effects, surface effects, crystal quality, electron irradiation, or scotch tape exfoliation (strain).

*Consideration of intrinsic effects for stacking disorder:*

Reported calculations of the $MoTe_2$ band structure suggest an intrinsic coupling between the layer stacking, total energy, and flake thickness. Specifically, Kim *et al* performed density functional theory (DFT) calculations on $MoTe_2$ and $WTe_2$, and examined the out-of-plane band structure (along $\Gamma$ to A)[34]. For $MoTe_2$, the out-of-plane bands near the Fermi level show a much greater dispersion compared to $WTe_2$, indicting a larger inter-layer coupling. Additionally, Kim *et al* found that the bands along $\Gamma$ to A are significantly altered through the 1T´ to $T_d$ transition for $MoTe_2$, but not for $WTe_2$. Taken together, these findings suggest that in thin flakes of $MoTe_2$, the out-of-plane bands will experience a thickness-effect, which will then modulate the layer stacking energetics. In contrast, for $WTe_2$, thickness-effects (and their influence on the stacking energetics) should be reduced. This reasoning is in line with our observations of disordered stacking in $MoTe_2$ flakes, and ordered $T_d$ stacking in $WTe_2$ flakes.

To explicitly test this hypothesis, we performed DFT calculations for 1T´ and $T_d$ stacking in MoTe$_2$ and WTe$_2$, for bulk crystals, as well as thin films of various thickness. We note that the layer stacking energy scale for MoTe$_2$ is ~1 meV / formula unit, which is approaching the uncertainty of DFT calculations. Moreover, the layer stacking energetics are determined by inter-layer vdW interactions, which are challenging to capture with DFT and depend sensitively upon the chosen vdW correction. Hence, care must be taken in analyzing such calculations, and we evaluate the results from three separate vdW corrections: Grimme-D3 (Grimme), rev-vdW-DF2 (rev), and rev+U[35–40].

Table 2 provides the relative energy differences between the $T_d$ and 1T´ stacking arrangements, ΔE, for the different thicknesses. The reported energies correspond to fully relaxed structures at 0 K. For bulk MoTe$_2$ and WTe$_2$, all of the tested vdW corrections predict that the $T_d$ structure is lower in energy, consistent with experimental data and prior DFT results[9,34]. However, for the thin films, the three vdW corrections are not in quantitative or even qualitative agreement. Thus, we cannot make any definitive claims regarding the stability of 1T´ or $T_d$ stacking for any specific thin film thickness, for either WTe$_2$ or MoTe$_2$. However, all of the vdW corrections predict variation in ΔE as a function of thickness. This trend suggests that the layer stacking energetics in MoTe$_2$ are thickness-dependent, and that for certain thicknesses, alternative stacking sequences may be stabilized over the bulk $T_d$ ground state. This mechanism provides a possible explanation for disordered stacking in MoTe$_2$.

**Table 2.** Computed energy difference between the $T_d$ stacking and 1T´ stacking. Bolded values correspond to structures where the 1T´ stacking is lower in energy than $T_d$ (where ΔE is positive)

|  | Thickness (# of layers) | ΔE = $E_{Td}$ – $E_{1T'}$ (meV / formula unit) | | |
|---|---|---|---|---|
|  |  | Grimmes | Rev | Rev+U |
| MoTe$_2$ | Bulk | -0.42 | -0.11 | -0.27 |
|  | 5 | -0.12 | -0.14 |  |
|  | 4 | **0.29** | -0.20 | **3.23** |
|  | 3 | -0.21 | -0.14 | -256 |
|  | 2 | -0.22 | **0.01** | **0.66** |
| WTe$_2$ | Bulk | -2.55 | -1.51 |  |
|  | 4 | **5.04** | **3.08** |  |
|  | 3 | -0.62 | -0.43 |  |
|  | 2 | -0.82 | -0.93 |  |

Alternatively, the disordered stacking in MoTe$_2$ may be an entropic – rather than an energetic – effect. When calculating the total free energy of a material, the phonon energies dictate the vibrational entropy, and lower energy phonons yield a lower free energy[41]. In bulk MoTe$_2$, the energy of the inter-layer shear phonon is significantly higher for the $T_d$ phase than the 1T´ phase (1.71 versus 1.55 meV)[42]. Hence, it was argued by Heikes *et al* that the MoTe$_2$ stacking transition is driven by the differing inter-layer phonon energies and their influence on the vibrational entropy[9]. In the context of our work, if reducing the MoTe$_2$ flake thickness alters the energy of the inter-layer phonon modes, then the relative free energy of 1T´ and $T_d$ stacking would be affected. Indeed, it is well-established that the inter-layer phonon modes in MoTe$_2$ are strongly thickness-dependent[19]. Thus, it is possible that our observations of disordered stacking in MoTe$_2$ are driven by a coupling between thickness, phonon energy, and entropy.

**Discussion and Conclusion**

Our observation of disordered stacking in thin MoTe$_2$ flakes – but ordered stacking in WTe$_2$ flakes – has interesting parallels with the bulk MoTe$_2$ and WTe$_2$ stacking transitions. Based on single crystal neutron diffraction measurements, the bulk MoTe$_2$ 1T´ to T$_d$ phase transition occurs through an intermediate disordered phase, characterized by pseudo-random layer stacking arrangements and diffuse scattering along *L*[31]. Upon warming, the T$_d$ to 1T´ transition occurs through a metastable T$_d$* phase, which can be described as ↓↓↑↑ stacking[28]. In contrast, for bulk WTe$_2$ single crystals, the 1T´ to T$_d$ phase transition is abrupt upon both heating and cooling, with no intermediate phases or thermal hysteresis[16]. Thus, even in bulk, there is a propensity for disordered and alternative stacking sequences in MoTe$_2$ which is absent in WTe$_2$.

For device applications, well ordered phases are usually desirable, hence, strategies to obtain well-ordered stacking in MoTe$_2$ should be explored. For thin flakes, an out-of-plane electric field will couple to the ferroelectric polarization of the T$_d$ phase and lower its energy. Thus, gate biasing may stabilize T$_d$ order[8]. Electronic doping, either chemical or electrostatic, should also be explored to obtain well-ordered MoTe$_2$ flakes, since doping is thought to influence the stacking energetics[34].

In conclusion, we studied the effect of thickness on the layer stacking of exfoliated MoTe$_2$ flakes through atomic-resolution STEM-HAADF imaging, *in situ* cryogenic TEM, Raman spectroscopy, electronic resistance measurements, and DFT calculations. We found that thin exfoliated flakes of MoTe$_2$ are not 1T´ or T$_d$, but rather possess disordered layer stacking. Our results raise important questions regarding the electronic structure, topology, and the charge transport mechanisms in exfoliated MoTe$_2$ flakes, as well as how thickness may influence layer stacking in other 2D materials which exhibit stacking-dependent functionality, *e.g.* magnetic 2D materials. This work also highlights the importance of atomic-scale analysis in determining the structure of 2D materials.

# References

[1]   A.L. Sharpe, E.J. Fox, A.W. Barnard, J. Finney, K. Watanabe, T. Taniguchi, M.A. Kastner, D. Goldhaber-Gordon, Science 365 (2019) 605–608.
[2]   Y. Cao, V. Fatemi, S. Fang, K. Watanabe, T. Taniguchi, E. Kaxiras, P. Jarillo-Herrero, Nature 556 (2018) 43–50.
[3]   Z. Song, Z. Wang, W. Shi, G. Li, C. Fang, B.A. Bernevig, Phys. Rev. Lett. 123 (2019) 036401.
[4]   K. Yasuda, X. Wang, K. Watanabe, T. Taniguchi, P. Jarillo-Herrero, Science 372 (2021) 1458–1462.
[5]   M. Vizner Stern, Y. Waschitz, W. Cao, I. Nevo, K. Watanabe, T. Taniguchi, E. Sela, M. Urbakh, O. Hod, M. Ben Shalom, Science 372 (2021) 1462–1466.
[6]   T. Song, Z. Fei, M. Yankowitz, Z. Lin, Q. Jiang, K. Hwangbo, Q. Zhang, B. Sun, T. Taniguchi, K. Watanabe, M.A. McGuire, D. Graf, T. Cao, J.-H. Chu, D.H. Cobden, C.R. Dean, D. Xiao, X. Xu, Nat. Mater. 18 (2019) 1298–1302.
[7]   W. Bao, L. Jing, J. Velasco, Y. Lee, G. Liu, D. Tran, B. Standley, M. Aykol, S.B. Cronin, D. Smirnov, M. Koshino, E. McCann, M. Bockrath, C.N. Lau, Nat. Phys. 7 (2011) 948–952.
[8]   Z. Fei, W. Zhao, T.A. Palomaki, B. Sun, M.K. Miller, Z. Zhao, J. Yan, X. Xu, D.H. Cobden, Nature 560 (2018) 336–339.
[9]   C. Heikes, I.-L. Liu, T. Metz, C. Eckberg, P. Neves, Y. Wu, L. Hung, P. Piccoli, H. Cao, J. Leao, J. Paglione, T. Yildirim, N.P. Butch, W. Ratcliff, Phys. Rev. Mater. 2 (2018) 074202.
[10]  E.J. Sie, C.M. Nyby, C.D. Pemmaraju, S.J. Park, X. Shen, J. Yang, M.C. Hoffmann, B.K. Ofori-Okai, R. Li, A.H. Reid, S. Weathersby, E. Mannebach, N. Finney, D. Rhodes, D. Chenet, A. Antony, L. Balicas, J. Hone, T.P. Devereaux, T.F. Heinz, X. Wang, A.M. Lindenberg, Nature 565 (2019) 61–66.
[11]  W.G. Dawson, D.W. Bullett, J. Phys. C Solid State Phys. 20 (1987) 6159–6174.
[12]  Y. Deng, X. Zhao, C. Zhu, P. Li, R. Duan, G. Liu, Z. Liu, ACS Nano 15 (2021) 12465–12474.
[13]  Z. Wang, D. Gresch, A.A. Soluyanov, W. Xie, S. Kushwaha, X. Dai, M. Troyer, R.J. Cava, B.A. Bernevig, Phys. Rev. Lett. 117 (2016) 056805.
[14]  Z. Wang, B.J. Wieder, J. Li, B. Yan, B.A. Bernevig, Phys. Rev. Lett. 123 (2019) 186401.
[15]  H.P. Hughes, R.H. Friend, J. Phys. C Solid State Phys. 11 (1978) L103–L105.
[16]  Y. Tao, J.A. Schneeloch, A.A. Aczel, D. Louca, Phys. Rev. B 102 (2020) 060103.
[17]  R. He, S. Zhong, H.H. Kim, G. Ye, Z. Ye, L. Winford, D. McHaffie, I. Rilak, F. Chen, X. Luo, Y. Sun, A.W. Tsen, Phys. Rev. B 97 (2018) 041410.
[18]  D.A. Rhodes, A. Jindal, N.F.Q. Yuan, Y. Jung, A. Antony, H. Wang, B. Kim, Y. Chiu, T. Taniguchi, K. Watanabe, K. Barmak, L. Balicas, C.R. Dean, X. Qian, L. Fu, A.N. Pasupathy, J. Hone, Nano Lett. 21 (2021) 2505–2511.
[19]  Y. Cheon, S.Y. Lim, K. Kim, H. Cheong, ACS Nano 15 (2021) 2962–2970.
[20]  S. Paul, S. Karak, M. Mandal, A. Ram, S. Marik, R.P. Singh, S. Saha, Phys. Rev. B 102 (2020) 054103.
[21]  B. Su, Y. Huang, Y.H. Hou, J. Li, R. Yang, Y. Ma, Y. Yang, G. Zhang, X. Zhou, J. Luo, Z. Chen, Adv. Sci. 9 (2022) 2101532.
[22]  C. Cao, X. Liu, X. Ren, X. Zeng, K. Zhang, D. Sun, S. Zhou, Y. Wu, Y. Li, J.-H. Chen, 2D Mater. 5 (2018) 044003.
[23]  W. Wang, S. Kim, M. Liu, F.A. Cevallos, R.J. Cava, N.P. Ong, Science 368 (2020) 534–537.
[24]  A. Tiwari, F. Chen, S. Zhong, E. Drueke, J. Koo, A. Kaczmarek, C. Xiao, J. Gao, X. Luo, Q. Niu, Y. Sun, B. Yan, L. Zhao, A.W. Tsen, Nat. Commun. 12 (2021) 2049.

[25] S. Lai, H. Liu, Z. Zhang, J. Zhao, X. Feng, N. Wang, C. Tang, Y. Liu, K.S. Novoselov, S.A. Yang, W. Gao, Nat. Nanotechnol. 16 (2021) 869–873.
[26] S.-Y. Chen, T. Goldstein, D. Venkataraman, A. Ramasubramaniam, J. Yan, Nano Lett. 16 (2016) 5852–5860.
[27] K. Zhang, C. Bao, Q. Gu, X. Ren, H. Zhang, K. Deng, Y. Wu, Y. Li, J. Feng, S. Zhou, Nat. Commun. 7 (2016) 13552.
[28] Y. Tao, J.A. Schneeloch, C. Duan, M. Matsuda, S.E. Dissanayake, A.A. Aczel, J.A. Fernandez-Baca, F. Ye, D. Louca, Phys. Rev. B 100 (2019) 100101.
[29] S. Cho, S.H. Kang, H.S. Yu, H.W. Kim, W. Ko, S.W. Hwang, W.H. Han, D.-H. Choe, Y.H. Jung, K.J. Chang, Y.H. Lee, H. Yang, S.W. Kim, 2D Mater. 4 (2017) 021030.
[30] M. Yankowitz, J.I.J. Wang, A.G. Birdwell, Y.-A. Chen, K. Watanabe, T. Taniguchi, P. Jacquod, P. San-Jose, P. Jarillo-Herrero, B.J. LeRoy, Nat. Mater. 13 (2014) 786–789.
[31] J.A. Schneeloch, C. Duan, J. Yang, J. Liu, X. Wang, D. Louca, Phys. Rev. B 99 (2019) 161105.
[32] F.-T. Huang, S. Joon Lim, S. Singh, J. Kim, L. Zhang, J.-W. Kim, M.-W. Chu, K.M. Rabe, D. Vanderbilt, S.-W. Cheong, Nat. Commun. 10 (2019) 4211.
[33] M. Kim, S. Han, J.H. Kim, J.-U. Lee, Z. Lee, H. Cheong, 2D Mater. 3 (2016) 034004.
[34] H.-J. Kim, S.-H. Kang, I. Hamada, Y.-W. Son, Phys. Rev. B 95 (2017) 180101.
[35] X. Gonze, C. Lee, Phys. Rev. B 55 (1997) 10355–10368.
[36] G. Kresse, D. Joubert, Phys. Rev. B 59 (1999) 1758–1775.
[37] G. Kresse, J. Furthmüller, Comput. Mater. Sci. 6 (1996) 15–50.
[38] P.E. Blöchl, Phys. Rev. B 50 (1994) 17953–17979.
[39] S. Grimme, J. Antony, S. Ehrlich, H. Krieg, J. Chem. Phys. 132 (2010) 154104.
[40] J.P. Perdew, K. Burke, M. Ernzerhof, Phys. Rev. Lett. 77 (1996) 3865–3868.
[41] S. Wippermann, W.G. Schmidt, Phys. Rev. Lett. 105 (2010) 126102.
[42] J.A. Schneeloch, Y. Tao, J.A. Fernandez-Baca, G. Xu, D. Louca, Phys. Rev. B 105 (2022) 014102.